\documentclass[conference]{IEEEtran}
\IEEEoverridecommandlockouts
\usepackage{cite}
\usepackage{amsmath,amssymb,amsfonts}
\usepackage{algorithmic}
\usepackage{graphicx}
\usepackage{textcomp}
\usepackage{xcolor}
\usepackage{svg}
\def\BibTeX{{\rm B\kern-.05em{\sc i\kern-.025em b}\kern-.08em
    T\kern-.1667em\lower.7ex\hbox{E}\kern-.125emX}}
\usepackage{wrapfig}
\usepackage{float}    
\usepackage{dblfloatfix}
\usepackage{bera}
\usepackage{listings,mdframed}
\usepackage{xcolor}
\usepackage{tikz}
\usepackage{pgfplots}
\usepackage{subcaption}
\pgfplotsset{compat=1.12}
\usepackage{pythonhighlight}
\definecolor{comment-text-color}{rgb}{0,0.8,0.6}
\lstdefinelanguage{json}{
    basicstyle=\normalfont\ttfamily\footnotesize,
    numbers=left,
    numberstyle=\scriptsize,
    stepnumber=1,
    numbersep=8pt,
    showstringspaces=false,
    breaklines=true,
    frame=lines,
    backgroundcolor=\color{background},
    literate=
     *{0}{{{\color{numb}0}}}{1}
      {1}{{{\color{numb}1}}}{1}
      {2}{{{\color{numb}2}}}{1}
      {3}{{{\color{numb}3}}}{1}
      {4}{{{\color{numb}4}}}{1}
      {5}{{{\color{numb}5}}}{1}
      {6}{{{\color{numb}6}}}{1}
      {7}{{{\color{numb}7}}}{1}
      {8}{{{\color{numb}8}}}{1}
      {9}{{{\color{numb}9}}}{1}
      {:}{{{\color{punct}{:}}}}{1}
      {,}{{{\color{punct}{,}}}}{1}
      {\{}{{{\color{delim}{\{}}}}{1}
      {\}}{{{\color{delim}{\}}}}}{1}
      {[}{{{\color{delim}{[}}}}{1}
      {]}{{{\color{delim}{]}}}}{1},
}

\lstdefinelanguage{llvm}{
	sensitive=true,
	alsoletter={\%},
	comment=[l]{;},
	string=[b]{"},
	keywords=[1]{
add, addrspacecast, alloca, and, ashr, atomicrmw, bitcast, br, call, cmpxchg,
extractelement, extractvalue, fadd, fcmp, fdiv, fence, fmul, fpext, fptosi,
fptoui, fptrunc, frem, fsub, getelementptr, icmp, indirectbr, insertelement,
insertvalue, inttoptr, invoke, landingpad, load, lshr, mul, or, phi, ptrtoint,
resume, ret, sdiv, select, sext, shl, shufflevector, sitofp, srem, store, sub,
switch, to, trunc, udiv, uitofp, unreachable, urem, va_arg, xor, zext
	},
	keywords=[2]{
acq_rel, acquire, addrspace, alias, align, alignstack, alwaysinline, any,
anyregcc, appending, arcp, arm_aapcs_vfpcc, arm_aapcscc, arm_apcscc, asm,
atomic, attributes, available_externally, blockaddress, builtin, byval, c,
catch, cc, ccc, cleanup, cold, coldcc, comdat, common, constant, datalayout,
declare, default, define, dereferenceable, dllexport, dllimport, eq, exact,
exactmatch, extern_weak, external, externally_initialized, false, fast, fastcc,
filter, gc, ghccc, global, hidden, inalloca, inbounds, initialexec, inlinehint,
inreg, intel_ocl_bicc, inteldialect, internal, jumptable, largest, linkonce,
linkonce_odr, localdynamic, localexec, max, min, minsize, module, monotonic,
msp430_intrcc, musttail, naked, nand, ne, nest, ninf, nnan, noalias, nobuiltin,
nocapture, noduplicate, noduplicates, noimplicitfloat, noinline, nonlazybind,
nonnull, noredzone, noreturn, nounwind, nsw, nsz, null, nuw, oeq, oge, ogt, ole,
olt, one, opaque, optnone, optsize, ord, personality, prefix, preserve_allcc,
preserve_mostcc, private, prologue, protected, ptx_device, ptx_kernel, readnone,
readonly, release, returned, returns_twice, samesize, sanitize_address,
sanitize_memory, sanitize_thread, section, seq_cst, sge, sgt, sideeffect,
signext, singlethread, sle, slt, spir_func, spir_kernel, sret, ssp, sspreq,
sspstrong, tail, target, thread_local, triple, true, type, ueq, uge, ugt, ule,
ult, umax, umin, undef, une, unnamed_addr, uno, unordered, unwind, uselistorder,
uselistorder_bb, uwtable, volatile, weak, weak_odr, webkit_jscc, x,
x86_64_sysvcc, x86_64_win64cc, x86_fastcallcc, x86_stdcallcc, x86_thiscallcc,
x86_vectorcallcc, xchg, zeroext, zeroinitializer
	},
	keywords=[3]{
i1, i2, i3, i4, i5, i6, i7, i8, i9, i10, i11, i12, i13, i14, i15, i16, i17, i18,
i19, i20, i21, i22, i23, i24, i25, i26, i27, i28, i29, i30, i31, i32, i33, i34,
i35, i36, i37, i38, i39, i40, i41, i42, i43, i44, i45, i46, i47, i48, i49, i50,
i51, i52, i53, i54, i55, i56, i57, i58, i59, i60, i61, i62, i63, i64, i80, i512,
void, half, float, double, fp128, x86_fp80, ppc_fp128, x86_mmx, label, metadata
	},
}

\lstdefinestyle{c}{
	commentstyle=\color{comment},
	stringstyle=\color{string},
	keywordstyle=\color{keyword},
	basicstyle=\footnotesize\ttfamily,
	numbers=none,
	numberstyle=\tiny,
	numbersep=5pt,
	frame=lines,
	breaklines=true,
	prebreak=\raisebox{0ex}[0ex][0ex]{\ensuremath{\hookleftarrow}},
	showstringspaces=false,
	upquote=true,
	tabsize=2,
}
\usepackage[english]{babel}
\usepackage{blindtext}
\usepackage{xcolor}
\usepackage{listings}
\usepackage{enumerate}
\usepackage{siunitx}
\usepackage{booktabs}
\usepackage{hyperref}
\hypersetup{
     colorlinks   = true,
     citecolor    = blue
}
\definecolor{red}{rgb}{1,0.,0}
\renewcommand{\floatpagefraction}{.98}%
\usepackage[compatibility=false]{caption}
\usepackage{tcolorbox}

\begin{document}
\renewcommand{\floatpagefraction}{.98}%
\title{A MLIR Dialect for Quantum Assembly Languages
\thanks{This manuscript has been authored by UT-Battelle, LLC under Contract No. DE-AC05-00OR22725 with the U.S. Department of Energy. The United States Government retains and the publisher, by accepting the article for publication, acknowledges that the United States Government retains a non-exclusive, paid-up, irrevocable, world-wide license to publish or reproduce the published form of this manuscript, or allow others to do so, for United States Government purposes. The Department of Energy will provide public access to these results of federally sponsored research in accordance with the DOE Public Access Plan. (http://energy.gov/downloads/doe-public-access-plan).}
}
\author{\IEEEauthorblockN{Alexander McCaskey\IEEEauthorrefmark{1}\IEEEauthorrefmark{2} and Thien Nguyen\IEEEauthorrefmark{1}\IEEEauthorrefmark{2}}
\IEEEauthorblockA{\IEEEauthorrefmark{1}Computer Science and Mathematics Division, Oak Ridge National Laboratory, Oak Ridge, TN, 37831, USA}
\IEEEauthorblockA{\IEEEauthorrefmark{2}Quantum Science Center, Oak Ridge National Laboratory, Oak Ridge, TN, 37831, USA}
}

\maketitle

%%%%%%%%%%%%%%%%%%%%%%%%%%
\begin{abstract}
We demonstrate the utility of the Multi-Level Intermediate Representation (MLIR) for quantum computing. Specifically, we extend MLIR with a new quantum dialect that enables the expression and compilation of common quantum assembly languages. The true utility of this dialect is in its ability to be lowered to the LLVM intermediate representation (IR) in a manner that is adherent to the quantum intermediate representation (QIR) specification recently proposed by Microsoft. We leverage a \texttt{qcor}-enabled implementation of the QIR quantum runtime API to enable a retargetable (quantum hardware agnostic) compiler workflow mapping quantum languages to hybrid quantum-classical binary executables and object code. We evaluate and demonstrate this novel compiler workflow with quantum programs written in OpenQASM 2.0. We provide concrete examples detailing the generation of MLIR from OpenQASM source files, the lowering process from MLIR to LLVM IR, and ultimately the generation of executable binaries targeting available quantum processors. 
\end{abstract}

\begin{IEEEkeywords}
quantum computing, quantum programming, quantum simulation, programming languages
\end{IEEEkeywords}

%%%%%%%%%%%%%%%%%%%%%%%%%
\section{Introduction}
%\fix{goals here - introduce disjointed nature of quantum programming tools - frameworks outputing qasm strings. quantum-classical implies need for integration with existing classical compiler work. need for quantum compilers. }
The availability of noisy quantum processing units (QPUs) from a variety of hardware vendors has raised new research and development questions into application use cases, programming model\cite{Dumitrescu2018,McCaskey2019,qsharp, scaffold}. It is anticipated that QPUs may serve as co-processors or accelerators within existing classical scientific computing workflows in a post-exascale computing era \cite{McCaskeyICRC2018}. The heterogeneous quantum-classical characteristic of this model of computation necessitates the development of quantum languages, compilers, and associated frameworks and tooling that tightly integrate with existing classical infrastructures. For example, the LLVM compiler infrastructure \cite{llvm} has proven critical to existing classical accelerator-based computing workflows, and we anticipate that building upon this infrastructure for quantum co-processing will precipitate required tight integration between classical and quantum compute resources. A number of efforts are leveraging LLVM and Clang to provide GPU-based programming models, compilers, and tools \cite{llvm-openmp,llvm-openacc}, and quantum computing may be well-suited to build upon these common ideas and concepts. 

\begin{figure*}[t!]
\centering
\includegraphics[width=.8\textwidth]{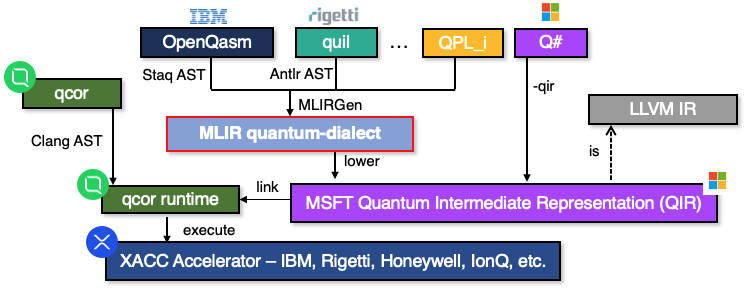}
\caption{Our goal is to enable progressive lowering of quantum assembly languages to executable binaries via the quantum intermediate representation (QIR). We provide a new dialect to MLIR for quantum assembly languages that can be lowered to LLVM IR adherent to the QIR specification. Linking to a \texttt{qcor}-enabled, QIR runtime library implementation will enables a retargetable compiler that produces executable binaries and object code for available quantum co-processors.}
\label{fig:qcor_layers}
\end{figure*}
Recently, a novel infrastructure for domain specific compilation has been put forward under the LLVM project umbrella. The Multi-Level Intermediate Representation (MLIR) represents a unique approach to the expression of compiler intermediate representations that promotes a variety of abstraction levels (close to the source language all the way down to machine level IR), enables progressive lowering from high-level IR abstractions to lower ones, and promotes easy and efficient extensibility and reusability of IR components \cite{mlir}. It has primarily been leveraged in classical heterogeneous computing workflows to enable common machine learning application compilation to available classical co-processors (GPUs, TPUs, FPGAs, etc.) \cite{mlir-nn, mlir-gpu}. As a domain-specific compiler infrastructure, the MLIR is be well suited for the development of compiler infrastructures mapping quantum assembly languages to executables or object code targeting available quantum co-processors. The development of a novel MLIR extension for quantum computing would enable an integration platform for vendor-specific languages and promote compilation in a quantum-backend-retargetable manner. 

In an effort to move quantum co-processing compiler tools towards integration with LLVM, researchers at Microsoft have recently introduced a novel specification called the quantum intermediate representation (QIR) \cite{qir}. This work defines a number of low-level LLVM callable IR instructions and opaque data types that are necessary for typical quantum computing workflows and tasks. The overall goal is to provide a unified representation that quantum languages and compilers can map to in an effort to promote reusability of common quantum compilation tools, strategies, and backend runtime implementations. The integration of the MLIR with the recently proposed QIR specification would directly enable quantum code optimization and progressive lowering to executable code in a quantum language and hardware agnostic fashion. Extending MLIR for quantum assembly languages would enable existing languages to quickly map to the LLVM IR, and then through appropriate library implementation linkage, execute on available quantum processors. This whole workflow would be directly integrated with popular classical compiler infrastructures and promote a hybrid quantum-classical workflow that is tightly integrated. 

In this work, we present a novel MLIR dialect for quantum assembly languages and implement lowering routines to map this representation to an instance of the LLVM IR adherent to the QIR specification. We demonstrate this workflow by providing a parser for the OpenQASM language that generates a corresponding MLIR instance. Ultimately we provide a compiler for quantum assembly languages (with OpenQASM as a first example) that takes quantum language source files and produces executables or library code that targets a desired quantum co-processor. Our work implements the QIR specification with a quantum runtime implementation backed by the \texttt{qcor} quantum-classical compiler platform, enabling execution on currently available quantum computers. 

%This work is structured as follows - first, we provide necessary background on the MLIR, QIR, and \texttt{qcor} in order to setup our presentation of an MLIR dialect for quantum computing. Next, we describe the MLIR Quantum Dialect, its structure, features, and goals. We then demonstrate our architecture for lowering quantum operations in this dialect to LLVM instructions adherent to the QIR specification, and detail a QIR runtime API implementation delegating to the QCOR backend QPU execution infrastructure. Finally, we evaluate the utility of this approach and provide illustrative tests and benchmarks for the OpenQASM 2.0 quantum language. 

\section{Background}
Here we provide some necessary background on the underlying technologies and specifications we leverage to provide a mechanism for compiling quantum languages through the MLIR, lowered to the LLVM IR, and ultimately to binary executables or object code. At a high-level, our approach can be seen in Figure \ref{fig:qcor_layers}. We put forward an extension to the MLIR infrastructure specifically for quantum computing, with the intention that any and all quantum languages are able to be parsed and mapped to an instance of this representation. We then provide the implementation necessary to map that representation to the LLVM IR which can then be linked with a backend-agnostic quantum runtime, thus enabling any quantum language to be compiled to an executable binary. Specifically, we detail at a high-level the internal structure of the MLIR, the new quantum intermediate representation (QIR) specification, and the QCOR quantum-classical compiler platform recently put foward at ORNL, which serves as the underlying quantum runtime enabling execution of the generated QIR code. 

\paragraph*{\textbf{MLIR}} The Multi-Level Intermediate Representation (MLIR) is a novel approach for constructing new domain specific compilers in a reusable and extensible manner. It puts forward a novel single-static assignment (SSA)-based intermediate representation for compiler developers that enables the encoding of higher levels of language abstraction, and the ability to progressively lower this representation to other levels of IR abstraction, ultimately targeting a machine-level IR (e.g. LLVM IR). For a full description of the framework, we invite the interested reader to see \cite{mlir, mlir-doc}. Here we detail a few key points that are important for the discussion of the work presented here. 

The key abstraction promoted by the MLIR is an \emph{operation}, represented in the framework as an \texttt{Op} class. Everything in the language (instructions, functions, composites of functions) is modeled as an \texttt{Op}, which is itself an extension point to the framework (developers can contribute new operations to define new semantics and functionality). Each \texttt{Op} has a unique string identifier (the operation name), can take and return zero or more values (the class \texttt{mlir::Value}), and retains a unique dictionary of constant information related to the \texttt{Op} called \texttt{Attributes}. Since these \texttt{Attributes} must be constant, they are used to represent compile-time information about the operation. 

Developers contribute new \texttt{Ops} via the introduction of a custom \texttt{Dialect} subclass. \texttt{Dialects} represent a logical grouping of \texttt{Ops} under a domain-specific and unique namespace string identifier. Each new \texttt{Dialect} registers at creation its provided set of operations, and in this way promotes modularity and extensibility of language parsing and lowering within the MLIR system. MLIR provides a builtin \texttt{Dialect} for expressing common concepts such as functions and modules (composites of functions). Other useful \texttt{Dialects} have been contributed as well - one representing the LLVM IR, others for GPU-accelerated programming models, and still others for operating on vectors and tensors common to machine-learning application workflows.

Critically, MLIR provides a robust infrastructure for transformations on operations that enable progressive lowering of the representation to lower levels of IR abstraction. The key concepts put forward here are the \texttt{ConversionTarget} and rewrite patterns represented as \texttt{ConversionPattern} class instances. First, one specifies a \texttt{ConversionTarget} sub-type that enforces the requirement that all transformed or re-written \texttt{Ops} are \emph{legal} with respect to that target \texttt{Dialect} (i.e. the \texttt{LLVMConversionTarget} enforces the requirement that all resultant operations are valid LLVM dialect operations). Next, for each \texttt{Op} in the custom \texttt{Dialect} that should be lowered, one can define a \texttt{ConversionPattern} sub-type that exposes a mechanism for specifying a replacement operation while marking the original operation as \emph{erased}. Ultimately, these extensions can be contributed to the MLIR framework and leveraged as part of a built-in \texttt{mlir::PassManager} pipeline that affects lowering of the high-level custom \texttt{Dialect} down to the conversion target. 

\paragraph*{\textbf{QIR}} The quantum intermediate representation (QIR) specification has recently been published and promoted by researchers at Microsoft and is intended to provide a common, unified compiled representation for quantum languages targeting any gate-based quantum computing platform. Critically, QIR is based on the LLVM IR (ultimately it is the LLVM IR) and specifies key concepts and rules for representing quantum instructions, register allocation, qubit addressing, and measurement retrieval. This approach promotes the utility of the full classical capabilities that LLVM provides, thereby enabling a mechanism for robust classical computation in tandem with quantum co-processing or acceleration. 

For the full specification, we direct the reader to \cite{qirspec}, however here we wish to highlight a few key points that are critical for the work described here. The key concepts in the specification are register allocation, qubit addressing, and instruction invocation, and we provide an example QIR code in Figure \ref{fig:qir_code} to demonstrates these concepts. First note that the specification leaves qubits as an opaque type so that implementations of an associated runtime library can define it in a way that best fits the implementation. Similarly, a measurement result type is also marked as opaque. Next, all runtime functions are declared but not implemented and are intended to be provided externally and linked in later. This design decision enables maximum flexibility to swap out various quantum runtime implementations for specific hardware backends. The specification defines a qubit register allocation function - \texttt{\_\_quantum\_\_rt\_\_qubit\_allocate\_array} - which takes as input the number of qubits to allocate and returns a pointer to the \texttt{Qubit} array. Individual qubits are queried or addressed via a declared \texttt{\_\_quantum\_\_rt\_\_array\_get\_element\_ptr\_1d} function, which takes an allocated qubit register and a qubit element index integer and returns a pointer to that \texttt{Qubit}. Quantum instructions are exposed as declared \texttt{\_\_quantum\_\_qis\_\_INSTNAME}, where \texttt{INSTNAME} is the name of the instruction (\texttt{h} for Hadamard, {cnot} for CNOT, etc.). The remainder of the code snippet follows any typical LLVM IR disassembled code (an LLVM *.ll file) where each line is an instruction call in the SSA form. A \texttt{call} is made to the quantum runtime functions with given input values and return results are stored to new value names. Figure \ref{fig:qir_code} allocates a register of size 2, gets the $0^{th}$ qubit, executes a Hadamard on that qubit to place it in superposition, gets the $1^{st}$ qubit, executes a CNOT to entangle the qubits, and finally measures both to produce single-bit \texttt{Results}. 
% \begin{center}%[20]{r}[\dimexpr.5\width+.5\columnsep\relax]{.75\textwidth}
    \begin{figure*}
    \lstset {language=llvm, linewidth=\textwidth}
    \begin{lstlisting}
%Array = type opaque
%Result = type opaque
%Qubit = type opaque

declare %Result* @__quantum__qis__mz(%Qubit* %0) 
declare void @__quantum__qis__cnot(%Qubit* %0, %Qubit* %1) 
declare void @__quantum__qis__h(%Qubit* %0) 

declare %Array* @__quantum__rt__qubit_allocate_array(i64 %0) 
declare void @__quantum__rt__qubit_release_array(%Array* %0) 
declare i8* @__quantum__rt__array_get_element_ptr_1d(%Array* %0, i64 %1) 

define i32 @main(i32 %0, i8** %1)  {
  %4 = call %Array* @__quantum__rt__qubit_allocate_array(i64 2)
  %5 = call i8* @__quantum__rt__array_get_element_ptr_1d(%Array* %4, i64 0)
  %6 = bitcast i8* %5 to %Qubit**
  %7 = load %Qubit*, %Qubit** %6, align 8
  call void @__quantum__qis__h(%Qubit* %7)
  %8 = call i8* @__quantum__rt__array_get_element_ptr_1d(%Array* %4, i64 1)
  %9 = bitcast i8* %8 to %Qubit**
  %10 = load %Qubit*, %Qubit** %9, align 8
  call void @__quantum__qis__cnot(%Qubit* %7, %Qubit* %10)
  %11 = call %Result* @__quantum__qis__mz(%Qubit* %7)
  %12 = call %Result* @__quantum__qis__mz(%Qubit* %10)
  call void @__quantum__rt__qubit_release_array(%Array* %4)
  ret i32 0
}
\end{lstlisting}
\captionof{figure}{Bell state measurement expressed in the QIR. Qubits and Results are opaque types, while quantum runtime functions are declared and implementations are expected to be linked in later.}
\label{fig:qir_code}
\end{figure*}
% \end{center}

These are only a few of the quantum runtime functions detailed by the QIR specification, but this should give an overall perspective on the structure of a compiled QIR program. Since this is just LLVM IR, we get standard program control flow structures and semantics for free, and we are free to integrate these quantum API calls with existing classical workflows. 
\begin{figure*}[t!]
\centering
\includegraphics[width=\textwidth]{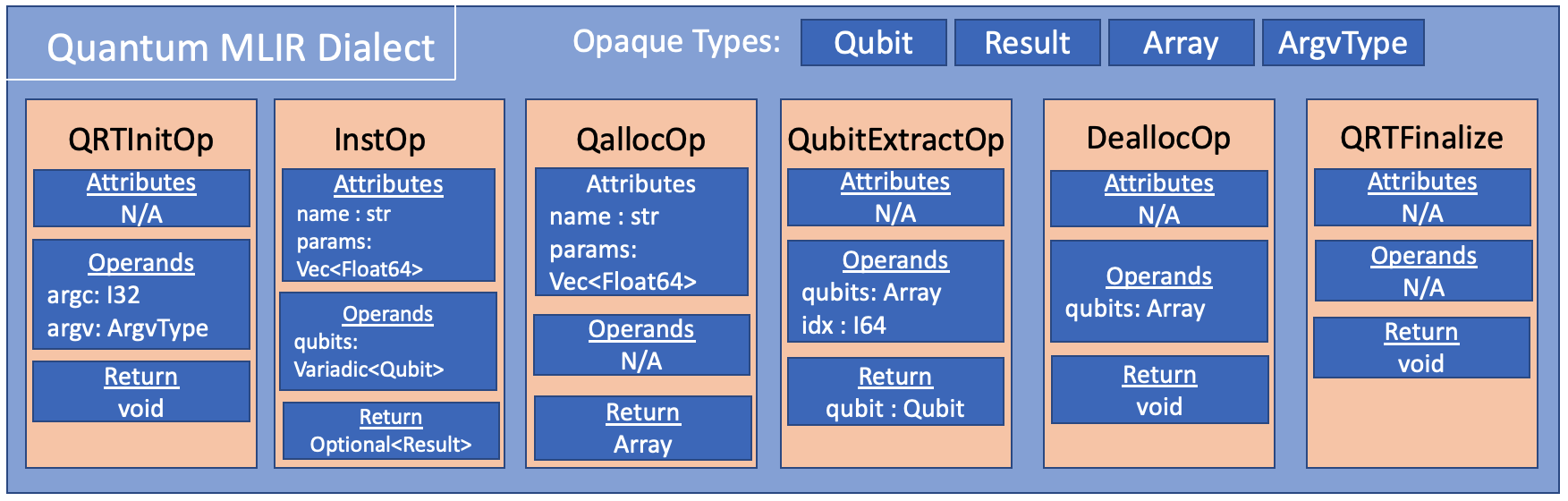}
\caption{Graphical representation of the operations contributed from the quantum MLIR dialect. These operations enable qubit array allocation and deallocation, qubit addressing, quantum instruction invocation, and a mechanism for initializing and finalizing the QIR runtime implementation. Each operation keeps track of constant attributes, and can take operands as input and return specific values.}
\label{fig:dialect_uml}
\end{figure*}

\paragraph*{\textbf{qcor}} Recently, researchers from Oak Ridge National Laboratory have put forward a novel language extension to C\texttt{++} for heterogeneous quantum-classical computing \cite{qcor}. The \texttt{qcor} compiler platform promotes a single-source programming model that enables programmers to write quantum functions natively in C\texttt{++} alongside classical code and compile to available quantum processors in a hardware-agnostic manner. \texttt{qcor} extends C\texttt{++} through a novel extension to the Clang plugin system \cite{clangsh} that enables the parsing of invalid quantum function bodies and their translation to valid C\texttt{++} API calls at compile time. To achieve compiler retargetability, \texttt{qcor} ultimately delegates to the XACC quantum programming framework \cite{mccaskey2020xacc}, which provides a system-level C\texttt{++} approach to quantum-classical programming, compilation, and execution in a service-oriented fashion. XACC exposes a number of interfaces that span the quantum compilation and execution workflow that enable quantum language parsing, quantum intermediate representation generation, circuit optimization, and backend execution on QPUs from IBM, Rigetti, D-Wave, Honeywell, among others. Ultimately, XACC provides \texttt{qcor} with an API for ubiquitous quantum compilation and execution tasks. 

To promote backend extensibility, \texttt{qcor} puts forward an extensible \texttt{QuantumRuntime} interface that exposes an API for quantum gate execution and can be implemented for both remotely hosted QPU protocols as well as future, tight integration models enabling CPU-QPU fast-feedback. These implementations directly delegate to the XACC \texttt{Accelerator} infrastructure to enable overall quantum compiler retargetability. For a complete background on the \texttt{qcor} compiler, we refer the reader to \cite{qcor, qcor-github, aideqc-docs}, but for the purposes of this work, the \texttt{QuantumRuntime} infrastructure is the primary background needed to detail the execution of MLIR-generated QIR code on available QPUs. 

\section{A Quantum MLIR Dialect}
To enable executable code generation targeting quantum co-processors from available quantum assembly languages, we seek to leverage the progressive lowering capabilities put forward by the MLIR compiler framework. This infrastructure provides a mechanism for mapping language-level intermediate representations down to classical assembly and object code via the LLVM IR. To properly interface quantum assembly languages with MLIR, we must provide a custom MLIR \texttt{Dialect} that exposes new operations (\texttt{Op} sub-classes) enabling the expression of common quantum assembly semantics. Specifically, we envision this work (this \texttt{Dialect} extension) as a first pass or initial prototype to demonstrate the overall utility of the MLIR for quantum code compilation and execution. Therefore, we keep our required feature-set small and define a \texttt{Dialect} that enables (1) quantum register allocation and deallocation, (2) qubit addressing within an allocated register, (3) simple quantum instruction invocation on addressed qubits, and (4) composition of quantum functions. These features enable one to perform simple circuit compilation and execution, and other higher-level languages should be able to map down to this lower-level of abstraction. 

We define a \texttt{QuantumDialect} subclass of \texttt{mlir::Dialect} which serves as an aggregator for custom \texttt{Ops} exposing these four quantum assembly features. Figure \ref{fig:dialect_uml} provides a graphical representation of this dialect and the operations it provides. The \texttt{QuantumDialect} exposes a unique namespace identifier - \texttt{quantum} - and registers six \texttt{Op} subclasses under the \texttt{quantum} namespace: the \texttt{QRTInitOp}, \texttt{QRTFinalizeOp}, \texttt{QallocOp}, \texttt{DeallocOp}, \texttt{QubitExtractOp}, and \texttt{InstOp}. The \texttt{QRTInitOp} and \texttt{QRTFinalizeOp} enable the initialization and finalization of a provided QIR runtime library implementation. Specifically, \texttt{QRTInitOp} expects the typical \texttt{argc}, \texttt{argv} input arguments coming from \texttt{main()}. We define an \texttt{mlir::OpaqueType} to describe the incoming \texttt{char**} argument.

The \texttt{QallocOp} performs qubit register allocation and exposes a unique name - \texttt{qalloc} - as well as a dictionary of compile-time attributes indicating the size and unique variable name for the register. It returns an \texttt{Array}, and opaque type that maps directly to the QIR array concept. The \texttt{Array} is assumed to be an array of the \texttt{Qubit} \texttt{OpaqueType}. The \texttt{QallocOp} does not take any input argument operands since we treat the register size as a compile-time constant, however, future iterations may wish to treat the register size as an input operand. To address individual qubits, the quantum dialect exposes a \texttt{QubitExtractOp} which takes and input \texttt{Array} value and a 64-bit integer describing the index of the array to retrieve. It returns an instance of \texttt{Qubit}. In order to ensure allocated qubit registers are deallocated upon going out of scope, the dialect provides a \texttt{DeallocOp} which takes as input the \texttt{Array} of \texttt{Qubits} to deallocate. 
% \begin{center}%[20]{r}[\dimexpr.5\width+.5\columnsep\relax]{.75\textwidth}
    \begin{figure*}[t]
    \lstset {language=llvm}
    \begin{lstlisting}
module  {
  func @main(%arg0: i32, %arg1: !quantum.ArgvType) -> i32 {
    "quantum.init"(%arg0, %arg1) : (i32, !quantum.ArgvType) -> ()
    %0 = "quantum.qalloc"() {name = "q", size = 1 : i64} : () -> !quantum.Array
    %c0_i64 = constant 0 : i64
    %1 = "quantum.qextract"(%0, %c0_i64) : (!quantum.Array, i64) -> !quantum.Qubit
    %2 = "quantum.inst"(%1) {name = "h"} : (!quantum.Qubit) -> none
    "quantum.dealloc"(%0) : (!quantum.Array) -> ()
    "quantum.finalize"() : () -> ()
    %c0_i32 = constant 0 : i32
    return %c0_i32 : i32
  }
}
\end{lstlisting}
\captionof{figure}{Example MLIR instance using the quantum dialect. This code allocates a register of size 1, gets reference to that single qubit, executes a hadamard gate, and then deallocates the register.}
\label{fig:example_mlir}
\end{figure*}
% \end{center}

Finally, we define the \texttt{InstOp} to model any operation that affects execution of a specific quantum instruction on a provided qubit or qubits. This \texttt{Op} exposes the unique \texttt{inst} name and a compile-time attribute dictionary indicating the name of the instruction and any potential \texttt{double} parameters (gate rotation angles, for example). Once again, for this first prototype implementation, we treat potential parameters as compile-time constants, but future iterations may choose to move this to input operands. \texttt{Qubits} that the instruction operates on are treated as input operands, and are modeled as a variadic list of type \texttt{Qubit}, allowing one to invoke single or multi-qubit operations. \texttt{InstOps} return an optional \texttt{Result} - most instructions do not return anything, while measurement instruction calls will return a binary result (qubit measured 0 or 1). 

Figure \ref{fig:example_mlir} gives a simple example of a printed MLIR instance leveraging operations from the quantum dialect. This example affects the application of a hadamard gate on a single qubit. The example demonstrates the utility of the various operations and shows runtime initialization and finalization, quantum register allocation and deallocation, and qubit addressing and gate invocation.

\section{MLIR Generation}
\label{sec:mlirgen}
Now that we have discussed the overall architecture of the new MLIR quantum dialect, we turn our attention to an extensible mechanism for the generation of this representation. The goal here is to enable a unique extension point that language developers can implement to take source strings to an instance of the \texttt{mlir::ModuleOp} containing one or many \texttt{mlir::FuncOp} instances each composed of quantum dialect operations. 
    \begin{figure*}
    \lstset {language=C++}
    \begin{lstlisting}
class QuantumMLIRGenerator {
 public:
  QuantumMLIRGenerator(mlir::MLIRContext& ctx);
  virtual void initialize_mlirgen(bool add_entry_point = true,
                                  const std::string file_name = "") = 0;
  virtual void mlirgen(const std::string& src) = 0;
  mlir::OwningModuleRef get_module();
  virtual void finalize_mlirgen() = 0;
};
class OpenQasmMLIRGenerator : public QuantumMLIRGenerator {...}
class QuilMLIRGenerator : public QuantumMLIRGenerator {...}
class JaqalMLIRGenerator : public QuantumMLIRGenerator {...}
...
\end{lstlisting}
\captionof{figure}{Class definition for the Quantum MLIR generation extension point. Developers can sub-type this class to provide a mechanism to map language-specific source strings to an instance of the MLIR leveraging the quantum dialect.}
\label{fig:mlirgen}
\end{figure*}

We put forward the \texttt{QuantumMLIRGenerator} interface in Figure \ref{fig:mlirgen} and intend it to be implemented by quantum assembly language developers to take in a language source string and output an MLIR module. The interface is simple, but enables sub-types to perform any initialization, language parsing to tree-like representations, tree-to-MLIR translation, and any finalization routines. Clients of this interface can initialize the generator and indicate whether to produce an MLIR instance with a \texttt{main()} function entry point. This feature is useful for the purposes of compiling assembly representations to object code versus executables. The \texttt{mlirgen()} method is meant to be implemented by sub-types to generate language specific abstract syntax trees and map that tree to the corresponding MLIR quantum dialect operations. We demonstrate the implementation of this interface for OpenQASM in Section \ref{sec:eval}, and leave implementations for other languages like Quil \cite{quil} and Jaqal \cite{jaqal} for future work. 

\section{Lowering to QIR}
With a valid MLIR dialect for quantum assembly languages in place, we now turn our attention to the specific mechanisms for lowering quantum operations to valid LLVM dialect operations. Specifically, we seek a mechanism to lower to the LLVM IR in a manner that is adherent to the recently defined quantum intermediate representation (QIR). Fortunately, MLIR currently has a built-in dialect for the LLVM IR and puts forward a robust and general mechanism for translating operations from one dialect to another. We leverage this functionality, and provide translation mechanisms that map our custom \texttt{Op} data structures to valid operations in the LLVM MLIR Dialect in a way that is QIR specification adherent. 

The basic workflow for lowering one MLIR Dialect to the LLVM Dialect is demonstrated in Figure \ref{fig:passmanager}. MLIR provides a \texttt{mlir::PassManager} data structure that orchestrates the execution of custom and default passes that act to transform the input \texttt{mlir::ModuleOp}. Developers can contribute new passes as subclasses of the \texttt{mlir::PassWrapper} template type, here we provide one called \texttt{QuantumToLLVMLoweringPass}. This pass sets the conversion target to the LLVM dialect and contribute a custom \texttt{Op} lowering pattern for one of our quantum dialect operations. Once the pass manager workflow is run, the resultant \texttt{ModuleOp} should entirely consist of \texttt{Ops} from the LLVM Dialect. Finally, MLIR provides an API function for translating that \texttt{ModuleOp} to an LLVM \texttt{Module}, which is the top-level abstraction for LLVM IR. From there, one can readily lower to object code for the target machine.
% \begin{center}%[20]{r}[\dimexpr.5\width+.5\columnsep\relax]{.75\textwidth}
    \begin{figure}[b!]
    \lstset {language=C++}
    \begin{lstlisting}
// Create the PassManager for lowering 
// to LLVM MLIR and run it
mlir::PassManager pm(&context);
auto q_to_llvm = 
 std::make_unique<QuantumToLLVMLoweringPass>();
pm.addPass(q_to_llvm);
auto module_op = module.getOperation();
if (mlir::failed(pm.run(module_op))) {
  std::cout << "Pass Manager Failed\n";
  return 1;
}

//! module_op is now totally in LLVM Dialect !

// Now lower MLIR to LLVM IR
llvm::LLVMContext llvmContext;
auto llvmModule = 
  mlir::translateModuleToLLVMIR(module, 
                              llvmContext);
\end{lstlisting}
\captionof{figure}{Code snippet demonstrating how one may apply the MLIR pass management system to lower representations in the quantum dialect to equivalent expressions using the LLVM dialect.}
\label{fig:passmanager}
\end{figure}
% \end{center}
    \begin{figure*}[t!]
    \lstset {language=llvm}
    \begin{lstlisting}
module  {
  llvm.func @__quantum__rt__qubit_release_array(!llvm.ptr<struct<"Array", opaque>>)
  llvm.func @__quantum__qis__h(!llvm.ptr<struct<"Qubit", opaque>>)
  llvm.func @__quantum__rt__array_get_element_ptr_1d(!llvm.ptr<struct<"Array", opaque>>, i64) \ 
                                                                           -> !llvm.ptr<i8>
  llvm.func @__quantum__rt__qubit_allocate_array(i64) -> !llvm.ptr<struct<"Array", opaque>>
  llvm.func @__quantum__rt__finalize()
  llvm.func @__quantum__rt__initialize(i32, !llvm.ptr<ptr<i8>>) -> i32
  llvm.func @main(%arg0: i32, %arg1: !llvm.ptr<ptr<i8>>) -> i32 {
    %0 = llvm.call @__quantum__rt__initialize(%arg0, %arg1) : (i32, !llvm.ptr<ptr<i8>>) -> i32
    %0 = llvm.mlir.constant(1 : i64) : i64
    %1 = llvm.call @__quantum__rt__qubit_allocate_array(%0) : (i64) \
                                                        -> !llvm.ptr<struct<"Array", opaque>>
    %2 = llvm.mlir.constant(0 : i64) : i64
    %3 = llvm.call @__quantum__rt__array_get_element_ptr_1d(%1, %2) : \
                                    (!llvm.ptr<struct<"Array", opaque>>, i64) -> !llvm.ptr<i8>
    %4 = llvm.bitcast %3 : !llvm.ptr<i8> to !llvm.ptr<ptr<struct<"Qubit", opaque>>>
    %5 = llvm.load %4 : !llvm.ptr<ptr<struct<"Qubit", opaque>>>
    %6 = llvm.call @__quantum__qis__h(%5) : (!llvm.ptr<struct<"Qubit", opaque>>) -> !llvm.void
    %7 = llvm.call @__quantum__rt__qubit_release_array(%1) : \ 
                                            (!llvm.ptr<struct<"Array", opaque>>) -> !llvm.void
    %8 = llvm.call @__quantum__rt__finalize() : () -> !llvm.void
    %9 = llvm.mlir.constant(0 : i32) : i32
    llvm.return %9 : i32
  }
}
\end{lstlisting}
\captionof{figure}{Translated MLIR code in the LLVM dialect corresponding to the quantum dialect code in Figure \ref{fig:example_mlir}.}
\label{fig:llvm_dialect}
\end{figure*}

The goal of each conversion operation is to erase the \texttt{Op} from the quantum dialect and replace it with an \texttt{Op} from LLVM Dialect. Specifically, we replace each with its corresponding QIR runtime function call represented as a \texttt{mlir::CallOp} with input operand represented as a \texttt{LLVM::FuncOp}. Figure \ref{fig:llvm_dialect} demonstrates the same code in Figure \ref{fig:example_mlir} lowered to the LLVM MLIR dialect via our contributed lowering operations. For the \texttt{QRTInitOp} we provide a rewrite pattern that erases the operation and introduces a declaration and call to the function \texttt{\_\_quantum\_\_rt\_\_initialize()} and set its input operands to the block arguments coming from the parent \texttt{main()} function. For the \texttt{QallocOp}, we extract the size attribute and create a new LLVM \texttt{ConstantOp} to return a 64-bit integer representation of the size, which then serves as the input operand of a call to \texttt{\_\_quantum\_\_rt\_\_qubit\_allocate\_array()}. The return value for this new LLVM operation is an LLVM \texttt{PointerType} to the opaque \texttt{Array} structure. Extracting a qubit from that array, the \texttt{QubitExtractOp}, is rewritten as a call to \texttt{\_\_quantum\_\_rt\_\_array\_get\_element\_ptr\_1d()} with the allocated array pointer serving as the input operand. This function (as per the QIR specification) returns an \texttt{i8} pointer, so we therefore also introduce LLVM \texttt{BitcastOp} and \texttt{LoadOp} instances to map it to a qubit pointer (the opaque \texttt{Qubit} type). \texttt{InstOp} is mapped to a \texttt{\_\_quantum\_\_qis\_\_INSTNAME} where \texttt{INSTNAME} is extracted from the instruction operation's attributes dictionary. The input operand for this LLVM function operation is the casted qubit pointer. Finally, the \texttt{DeallocOp} is mapped to a call to \texttt{\_\_quantum\_\_rt\_\_qubit\_release\_array()} with input operand pointing to the original allocated qubit array value. We end with a call to \texttt{\_\_quantum\_\_rt\_\_finalize()} and add a return operation to terminate the block. 
These conversion operations are all contributed to the MLIR pass management system via our custom \texttt{QuantumToLLVMLoweringPass}. 

Once the quantum assembly code is represented using the LLVM MLIR dialect, we are able to lower that representation even further to the LLVM IR using the built-in \texttt{mlir::translateModuleToLLVMIR()} function. This returns an instance of the \texttt{llvm::Module} which we are then able to persist to a human-readable, LLVM assembly language \texttt{*.ll} file. We can then use existing tools to lower this even further to object code. Specifically, LLVM provides the \texttt{llvm-as} and \texttt{llc} tools which can lower this file to machine-specific object code. We can then link this object file with a custom QIR runtime API implementation to output a binary executable that will affect execution of the compiled quantum assembly code on a target quantum co-processor.

\section{QIR Runtime Implementation}
We provide an implementation of the QIR runtime API that is backed by the \texttt{qcor} quantum-classical compiler platform. This implementation stipulates that the \texttt{Array} opaque type maps to a \texttt{std::vector<int8\_t*>}, and the \texttt{Qubit} and \texttt{Result} map to a \texttt{uint64\_t}. We implement \texttt{\_\_quantum\_\_rt\_\_initialize()} to take in any input command line arguments and initialize the \texttt{qcor} runtime. This initialization sets the target quantum co-processor, and any related options, and selects the runtime mode. This mode can be one of two choices - the fault-tolerant, tightly integrated \texttt{ftqc} mode, or the remotely-hosted, loosely-coupled \texttt{nisq} mode (for currently available quantum processors). The mode can be configured from the command line as this initialization runtime function reads input command line arguments. 

The \texttt{qcor} quantum runtime exposes API calls for executing specific instructions on the backend quantum co-processor. For each implementation of \texttt{\_\_quantum\_\_qis\_\_INSTNAME}, we delegate execution to the appropriate \texttt{qcor} runtime API call (e.g. \texttt{\_\_quantum\_\_qis\_\_h(Qubit* q)} maps to \texttt{::quantum::h(qubit\& q)}). \texttt{qcor} represents quantum registers using a \texttt{qreg} data type that delegates to the \texttt{xacc::AcceleratorBuffer} concept. We implement \texttt{\_\_quantum\_\_rt\_\_qubit\_allocate\_array()} to create this underlying \texttt{AcceleratorBuffer} of the given register size. Calls to measure persist bit configurations to this allocated buffer, and return the measured bit configuration in the case of \texttt{FTQC} execution. 

Internally, \texttt{qcor} circuit execution builds upon the \texttt{xacc::Accelerator} subsystem. The \texttt{Accelerator} interface exposes a mechanism for execution of the XACC quantum intermediate representation as well as methods for querying backend-specific information like qubit connectivity, noise parameters, gate set, etc. XACC provides implementations of this interface for targeting all IBM quantum computing backends, the Rigetti QCS platform, Honeywell ion trap quantum computers, and a number of simulators that scale on Summit-like~\cite{summit} architectures. Our QIR runtime implementation inherits this extensibility and retargetability by delegating to the \texttt{qcor} runtime execution infrastructure. Assembly languages lowered to the QIR are therefore able to be lowered to executable binaries that target a wide range of quantum co-processors. 

\section{Evaluation}
\label{sec:eval}
Here we evaluate utility and efficacy of this quantum compiler infrastructure built upon the MLIR framework. We demonstrate its utility for the OpenQASM quantum assembly language, and describe integration efforts for compiled OpenQASM libraries with exiting \texttt{qcor} C\texttt{++} application codes. Our work provides a mechanism for compiling libraries of quantum codes written in OpenQASM that can then be included and linked to in future \texttt{qcor} programs. We end with a discussion and demonstration of the runtime of this language-to-object code workflow, and compare with the compiler workflow provided by Q\#. 

\subsection{OpenQASM Compiler}
OpenQASM is a popular assembly language for currently available IBM quantum processors \cite{openqasm}, is supported by a large number of software toolchains, and exposes language features like register allocation, qubit addressing, and instruction invocation, among others. For this work, we leverage an off-the-shelf parser for OpenQASM provided by the \texttt{staq} project \cite{staq}. This project provides a Clang-inspired OpenQASM parser in C\texttt{++} that translates language strings to a custom abstract syntax tree (AST) representation. Moreover, the project provides a robust visitor pattern on top of this representation that makes \texttt{staq} well-suited for key circuit transformation, optimization, and analysis tasks. We leverage this visitor pattern to walk the OpenQASM AST and map key nodes to corresponding MLIR \texttt{Ops} from our quantum dialect. 
\begin{figure}[t!] 
  \lstset {language=C++}
  \begin{lstlisting}
OpenQASM 2.0;
include "qelib1.inc";
qreg q[2];
creg c[2];             |
h q[0];                | staq ast-gen
cx q[0], q[1];         V
measure q -> c;

  \end{lstlisting}
  \begin{lstlisting}
|- Program
  ... qelib.inc (not included for brevity)...
  |- Register Decl(q[2], quantum)
  |- Register Decl(c[2])
  |- Declared(h)
    |- Var(q[0])
  |- Declared(cx)       key staq ast nodes:
    |- Var(q[0])           - RegisterDecl 
    |- Var(q[1])           - DeclaredGate
  |- Measure               - CNOTGate
    |- Var(q[0])           - UGate
    |- Var(c[0])           - MeasureStmt
  |- Measure
    |- Var(q[1])
    |- Var(c[1])
\end{lstlisting}
\caption{(top) A simple OpenQASM code to create a 2-qubit bell state followed by a measurement of all qubits. (bottom) \texttt{staq} AST representation demonstrating key nodes that need to be visited to map to an equivalent MLIR representation.}
\label{fig:oqasmtostaq}
\end{figure}

The \texttt{staq} AST exposes a top-level \texttt{ASTNode} type that all node-types derive from. The primary nodes that we care about in the generation of a \texttt{ModuleOp} composed of operations from the quantum dialect are the \texttt{RegisterDecl}, \texttt{UGate}, \texttt{CNOTGate}, \texttt{DeclaredGate}, and \texttt{MeasureStmt}. We provide a visitor implementation (\texttt{OpenQasmMLIRGenerator} that additionally sub-types the \texttt{QuantumMLIRGenerator} interface (see Section \ref{sec:mlirgen}) and implements \texttt{mlirgen()} to walk the AST and visit these specific nodes to create corresponding operations from our quantum dialect. Statements like \texttt{qreg q[2]} ultimately map to a \texttt{RegisterDecl} in the \texttt{staq} AST and visitation of that node provides the register variable name and size information (\texttt{q} and 2 for \texttt{qreg q[2]}). With this information, our visit implementation can construct a new \texttt{QallocOp} and set its constant \texttt{name} and \texttt{size} attributes and insert it into an pre-initialized \texttt{ModuleOp} class member on the visitor class implementation. Note that downstream \texttt{InstOps} and \texttt{QubitExtractOps} will need reference to the return register from the \texttt{QallocOp} as its input operands, therefore, we store all \texttt{QallocOp} instances in a symbol table stored on as a visitor class member.% We implement \texttt{UGate} and \texttt{CNOTGate} in a similar fashion to \texttt{DeclaredGate}.

Next, for gates that are not \texttt{U} or \texttt{CNOT} (these are separated out in the tree since they represent the common basis gate set on the IBM processor), we visit the \texttt{DeclaredGate} node (\texttt{h, rx, cx}, etc. - these gates ultimately lower to \texttt{U} and \texttt{CNOT}). This node provides information about the name of gate instruction and the qubits being operated on - specifically their address in the pre-allocated quantum register (\texttt{h q[0]} gives us the name \texttt{h}, the register \texttt{q}, and the qubit index 0). Within this visit implementation we leverage this information to construct a new \texttt{QubitExtractOp} that takes the correct \texttt{QallocOp} return operand retrieved from the symbol table, and a \texttt{mlir::ConstantOp} for generating the constant qubit index integer. The return operand from the \texttt{QubitExtractOp} can then be used as an input operand on a constructed \texttt{InstOp}, which is also seeded with corresponding constant attributes for the name of the instruction and any constant gate rotation parameters. 

We integrate this workflow within the \texttt{qcor} compiler platform. We provide a \texttt{qcor-mlir-tool} executable that takes an OpenQASM file as input and outputs an LLVM IR assembly file. Optionally, users can pass \texttt{-emit=mlir}, \texttt{-emit=mlir-llvm}, or \texttt{-emit=llvm} to print out the MLIR quantum, MLIR LLVM, or LLVM IR representations to standard out, respectively. Users may also use this tool to compile to LLVM IR with or without a \texttt{main()} entry point (the \texttt{-no-entrypoint} argument causes the workflow to output the LLVM IR without a \texttt{main} function). Note this is useful in mapping OpenQASM source files to executables or libraries for linking later. We have updated the \texttt{qcor} compiler driver to delegate to this tool when passed OpenQASM source files. Users can compile and execute OpenQASM codes as in Figure \ref{fig:compile_oqasm}.
\begin{figure}[t!]
  \lstset {language=C++}
  \begin{lstlisting}
$ qcor bell.qasm
$ ./a.out -qrt nisq -qpu ibm:ibmq_paris
\end{lstlisting}
\caption{Compile the OpenQASM code in Figure \ref{fig:oqasmtostaq} (first line). Run the resultant executable targeting an IBM physical backend (second line).}
\label{fig:compile_oqasm}
\end{figure}

It is interesting to note that the same parsing process can be performed for other quantum assembly languages. We anticipate that existing parsing mechanisms for languages such as quil, jaqal, and others rely on some sort of parse tree representation that can be traversed with specific general actions being invoked for differing node types. We observe that a number of languages provide an extended Bachus-Naur form grammar specification that could easily generate parse tree data structures through tools and frameworks like Antlr~\cite{ANTLR}. Frameworks such as this also provide listener and visitor patterns which can be leveraged to create quantum \texttt{Ops} and effectively map the language to our MLIR quantum dialect. 

\begin{figure}[b!]
  \lstset {language=C++}
  \begin{lstlisting}
$ qcor -no-entrypoint bell.qasm
$ ls
   bell.o bell.qasm
----------------------------------------------
#include "qcor.hpp"

// Macro that maps to 
// extern "C" void bell(qreg);
include_qcor_qasm(bell)

int main() {
   auto q = qalloc(2);
   // Function from bell.o
   bell(q)
   for (auto [bit, count] : q.counts()) {
      print(bit, ":", count);
   }
   return 0;
}
----------------------------------------------
$ qcor bell.o test.cpp -o test.x
$ ./test.x -qrt nisq -shots 2048 -qpu aer
00 : 1025
11 : 1023
\end{lstlisting}
\caption{Code snippet demonstrating the compilation of an OpenQASM code and subsequent linking within a standard \texttt{qcor} C\texttt{++} code.}
\label{fig:qreg}
\end{figure}

\subsection{Linking to Compiled Assembly Code}
\label{sec:linking_to_compiled_asm_code}
The ability to compile assembly languages with \texttt{qcor} using the \texttt{-no-entrypoint} command line option enables one to produce libraries of quantum functions that one should be able to link to and invoke from existing C\texttt{++} application codes. To do this, one would need some means to retrieve execution results from the linked OpenQASM function. In \texttt{qcor}, execution results are persisted to an allocated \texttt{qreg} instance, so what we need is a mechanism for passing an allocated \texttt{qreg} to the quantum function provided by the compiled object file. To achieve this, we extend the quantum dialect with a \texttt{SetQregOp} operation that takes as in put a value of opaque type \texttt{qreg}. Runtime implementations are free to define the \texttt{qreg} and provide it at the link phase, and we map it to the qcor \texttt{qreg} data type. The \texttt{SetQregOp} operation lowers to a new runtime function we define in the QIR API specification \texttt{\_\_quantum\_\_rt\_\_set\_qreg(qreg*)}, and is injected at the start of the top-level, entry-point quantum MLIR function. Our implementation provides this function and takes the \texttt{qreg} for use in persisting measurement results. 

This functionality promotes the ability for programmers to define quantum code in stand-alone OpenQASM source files, compile them with \texttt{qcor} (delegating to our \texttt{qcor-mlir-tool} to generate LLVM IR) using the \texttt{-no-entrypoint} flag, and link them into other \texttt{qcor} codes for invocation of the pre-compiled quantum assembly code. Figure \ref{fig:qreg} demonstrates this workflow, where users compile OpenQASM to an object file, call a pre-defined macro \texttt{include\_qcor\_qasm()} to setup the external linkage, and then call the function like any other C\texttt{++} function. 

\subsection{Compiler Benchmarks}
In this evaluation, we compare the compilation time of the QCOR compiler using the MLIR-based compilation workflow vs. the code-rewrite via syntax handler plugin approach. Specifically, in the hierarchical MLIR compilation workflow, the compilation runtime involves QASM to MLIR to QIR lowering and LLVM IR compilation (via \texttt{llvm-as} and \texttt{llc -filetype=obj}. On the other hand, when using the syntax handler, we embed the QASM source into the \texttt{.cpp} source file as a quantum kernel and use QCOR to compile the mixed quantum-classical source. 

\begin{figure}[t!]
\centering
\includegraphics[width=\columnwidth]{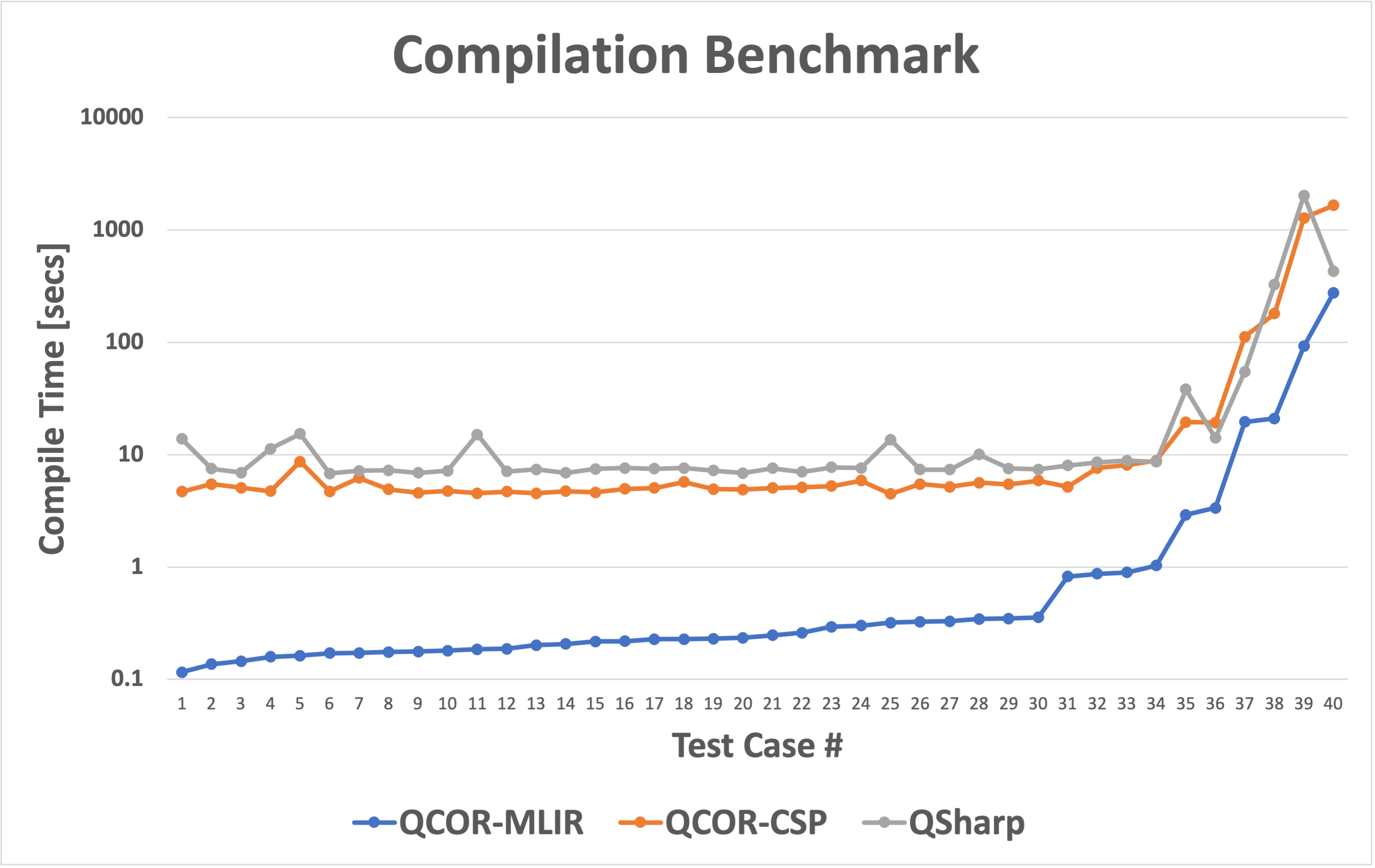}
\caption{Compilation time comparison between MLIR-based (QCOR-MLIR) and syntax handler plugin-based (QCOR-CSP) QCOR compilation and the Q\# compiler (QSharp). The test cases are flattened OpenQASM source files for the QCOR compiler and transpiled Q\# sources for the Q\# compiler. In this plot, the x-axis represents individual test cases sorted by the runtime of the QCOR-MLIR compiler.}
\label{fig:compile_time_compare}
\end{figure}

For comparison, we transpile the QASM test cases into Q\# and time the Q\# compilation\footnote{Command-line Q\# compiler (\texttt{qsc}) from Microsoft Quantum SDK version 0.14.2011120240 for Linux}. The result is shown in Figure~\ref{fig:compile_time_compare} for a set of OpenQASM source files~\cite{staq_benchmark_source} for common arithmetic quantum algorithms, such as adder, quantum Fourier transform, etc.  

The benefit of hierarchical lowering via MLIR to compilation time is significant (up to 1 to 2 order of magnitude) compared to the Clang syntax-hander plugin (CSP) approach~\cite{qcor}.  It is worth noting that the CSP approach comprises two steps: (1) transformation of embedded QASM into C++ data structures and (2) compilation of the rewritten source code. The second step is equivalent to the compilation workflow of the Q\# compiler since the QASM sources have been transpiled into Q\# code. 

Both the Q\# and QCOR-CSP compilers can handle higher-level language constructs, such as conditional statements, loops, function calls, etc., thus, are gearing towards quantum kernels written in that manner. On the other hand, direct QASM to MLIR then QIR lowering and external linkage, as demonstrated in sub-section~\ref{sec:linking_to_compiled_asm_code}, is a more efficient compilation approach to integrate pure QASM kernels into the heterogeneous quantum-classical programming workflow of QCOR.

\section{Conclusion}
We have presented a compilation infrastructure for quantum assembly languages that builds upon the extensible and reusable MLIR infrastructure and enables progressive lowering to the LLVM IR in a manner that is adherent to the Microsoft quantum intermediate representation (QIR). We extend the MLIR with a new dialect for quantum computing exposing operations for qubit memory management, addressing, and quantum instruction invocation. We have provided lowering routines for taking this operations to valid operations in the LLVM dialect. Our LLVM representation adheres to the QIR, and we provide a runtime API implementation that delegates execution to the \texttt{qcor} runtime library. We demonstrate the utility of this approach for a prototypical assembly language - OpenQASM. Our work enables quantum-classical compilation to executables or libraries, and for the latter can be leveraged within existing application use cases. 

We see this work as a foundational layer for continued development of quantum-classical compilers, frameworks, and tools. We envision the current OpenQASM workflow as a template that other developers may follow to extend this MLIR compiler workflow to other quantum assembly languages. This internal MLIR representation, and the lower QIR representation, can serve as a unified representation that enables integration across a number of quantum software research and development activities. We also note that a new version of OpenQASM (version 3.0) has recently been released that provides a more rich feature set enabling a hybrid quantum-classical approach at a higher-level of abstraction then low-level assembly. We belive this infrastructure could serve as a base foundation for the development of a true ahead-of-time compiler for OpenQASM 3.0 codes, enabling backend execution on available quantum co-processors.

\section*{Acknowledgment}
This material is based upon work supported by the U.S. Department of Energy, Office of Science, National Quantum Information Science Research Centers. ORNL is managed by UT-Battelle, LLC, for the US Department of Energy under contract no. DE-AC05-00OR22725.

\bibliographystyle{plain}
\bibliography{main}
\end{document}